\documentclass[12pt]{article}
\usepackage{epsfig}
\usepackage{graphicx}
\usepackage{lineno}
\usepackage{rotating}
\usepackage{lscape}
\usepackage{amsmath}
\usepackage[usenames]{color}
\usepackage{color}

\begin{document}
\begin{center}
\vspace*{1.0cm}

{\Large \bf{New experimental limits on double-beta decay of
osmium}}

\vskip 1.0cm

{\bf P.~Belli$^{a,b}$,
 R.~Bernabei$^{a,b,}$\footnote{Corresponding
author at: Dipartimento di Fisica, Universit\`{a} di Roma ``Tor
Vergata'', I-00133 Rome, Italy. E-mail address:
rita.bernabei@roma2.infn.it (R.~Bernabei)},
 F.~Cappella$^{c,d}$,
 V.~Caracciolo$^{a,b,e}$,
 R.~Cerulli$^{a,b}$,
 F.A.~Danevich$^{f}$,
 A.~Incicchitti$^{c,d}$,
 D.V.~Kasperovych$^{f}$,
 V.V.~Kobychev$^{f}$,
 G.P.~Kovtun$^{g,h}$,
 N.G.~Kovtun$^{g}$,
 M.~Laubenstein$^{e}$,
 V. Merlo$^{a,b,}$,
 D.V.~Poda$^{i}$,
 O.G.~Polischuk$^{f}$,
 A.P.~Shcherban$^{g}$,
 S.~Tessalina$^{j}$,
 V.I.~Tretyak$^{f}$
 }

\vskip 0.3cm

$^{a}${\it INFN, sezione di Roma ``Tor Vergata'', I-00133 Rome,
Italy}

$^{b}${\it  Dipartimento di Fisica, Universit\`{a} di Roma ``Tor
Vergata, I-00133 Rome, Italy''}

$^{c}${\it  INFN, sezione di Roma, I-00185 Rome,
Italy}

$^{d}${\it  Dipartimento di Fisica, Universit\`{a} di Roma ``La
Sapienza'', I-00185 Rome, Italy}

$^{e}${\it  INFN, Laboratori Nazionali del Gran Sasso, 67100
Assergi (AQ), Italy}

$^{f}${\it  Institute for Nuclear Research of NASU, 03028 Kyiv,
Ukraine}

$^{g}${\it  National Science Center ``Kharkiv Institute of Physics
and Technology'', 61108 Kharkiv, Ukraine}

$^{h}${\it V.N.~Karazin Kharkiv National University, 4, 61022
Kharkiv, Ukraine}

$^{i}${\it Universit\'{e} Paris-Saclay, CNRS/IN2P3, IJCLab, 91405
Orsay, France}

$^{j}${\it John de Laeter Centre for Isotope Research, GPO Box U
1987, Curtin University, Bentley, WA, Australia}

\end{center}

\vskip 0.5cm
\begin{abstract}
Double-beta processes in $^{184}$Os and $^{192}$Os were searched
for over 15851 h at the Gran Sasso National Laboratory (LNGS) of
the I.N.F.N. by using a 118 g ultra-pure osmium sample installed
on the endcap of a 112 cm$^3$ ultra-low-background broad-energy
germanium detector. New limits on double-electron capture and
electron capture with positron emission in $^{184}$Os were set at
the level of $\lim T_{1/2} \sim 10^{16}-10^{17}$ yr. In particular
the $2\nu$2K and $2\nu$KL decays of $^{184}$Os to the ground state
of $^{184}$W are restricted as $T_{1/2}\geq3.0\times 10^{16}$ yr
and $T_{1/2}\geq2.0\times 10^{16}$ yr, respectively. A lower limit
on the half-life for the double-beta decay of $^{192}$Os to the
first excited level of $^{192}$Pt was set as $\lim
T_{1/2}=2.0\times 10^{20}$ yr at 90\% C.L.

\end{abstract}

\vskip 0.2cm

Keywords: $^{184}$Os, $^{192}$Os, Double-beta decay,
Low-background gamma spectrometry

\section{Introduction}

Double-beta ($2\beta$) decay of atomic nuclei is a key process to
study properties of neutrino and weak interaction, and to search
for effects beyond the Standard Model of particles and
interactions (SM). While the $2\beta$ decay with the emission of two
neutrinos ($2\nu2\beta$) is allowed by the SM and already observed
in several nuclides with half-lives in the $(10^{18}-10^{24})$ yr
range \cite{Barabash:2020}, the neutrinoless mode of the decay
($0\nu2\beta$) is forbidden in the framework of the SM since the
process breaks the lepton number $L$ by two units
\cite{Deppisch:2012,Bilenky:2015,Delloro:2016,Vergados:2016,Dolinski:2019}
and probes the Majorana nature of the neutrino (the particle is equal
to its antiparticle) \cite{Majorana:1937,Schechter:1982}. The
$0\nu2\beta$ decay remains unobserved despite more than seventy
years of experimental attempts. The most sensitive experiments
give limits on the $0\nu2\beta$ decay half-life for several nuclei
at the level of $\lim T_{1/2}\sim(10^{24}-10^{26})$ yr
\cite{Agostini:2020,Azzolini:2019,Armengaud:2021,Adams:2020,Gando:2016}.
Assuming the mass mechanism of the $0\nu2\beta$ decay by exchange
of a virtual light Majorana neutrino, the region of half-life
limits corresponds to effective Majorana neutrino mass limits in
the range of $\lim \langle m_{2\beta}\rangle \sim (0.1-0.5)$ eV.
The range of $\langle m_{2\beta}\rangle$ limits is due to the
possible quenching of the axial vector coupling constant $g_A$
(used to calculate the effective neutrino mass from the
experimental half-life limits) \cite{Barea:2013,Suhonen:2019} and
the uncertainty of nuclear-matrix-element calculations (see, e.g.,
discussion and references in \cite{Giuliani:2018}).

The experimental sensitivity to double-electron capture (2EC),
electron capture with positron emission (EC$\beta^+$), and
double-positron decay ($2\beta^+$) is substantially lower than the
sensitivity of the experiments to search for $2\beta$ decay with
electrons emission ($2\beta^-$). At the same time, the mechanisms
of the neutrinoless 2EC, EC$\beta^+$ and $2\beta^+$ processes are
the same as for the $0\nu2\beta^-$ decay and thus, the investigations
of the 2EC, EC$\beta^+$ and $2\beta^+$ decays can give essentially
the same information about properties of the neutrino and the weak
interaction.

However, there are important arguments to develop experimental
methods to search for the $0\nu$2EC, $0\nu$EC$\beta^+$ and
$0\nu2\beta^+$ decays owing to the potential clarification of the
mechanism of the $0\nu2\beta^-$ decay when
observed: whether it is mediated by the mass
mechanism with exchange by virtual light Majorana neutrinos or
with a possible contribution of right-handed currents in the weak
interaction \cite{Hirsch:1994}. In addition there is an
interesting possibility of a resonant $0\nu$2EC process that can
increase the $0\nu$2EC decay probability up to the six orders of
magnitude in case the parent and daughter atoms' energies are degenerate
~\cite{Winter:1955a,Voloshin:1982,Bernabeu:1983,Krivoruchenko:2011,Blaum:2020}.

\begin{figure}[!ht]
 \begin{center}
 \mbox{\epsfig{figure=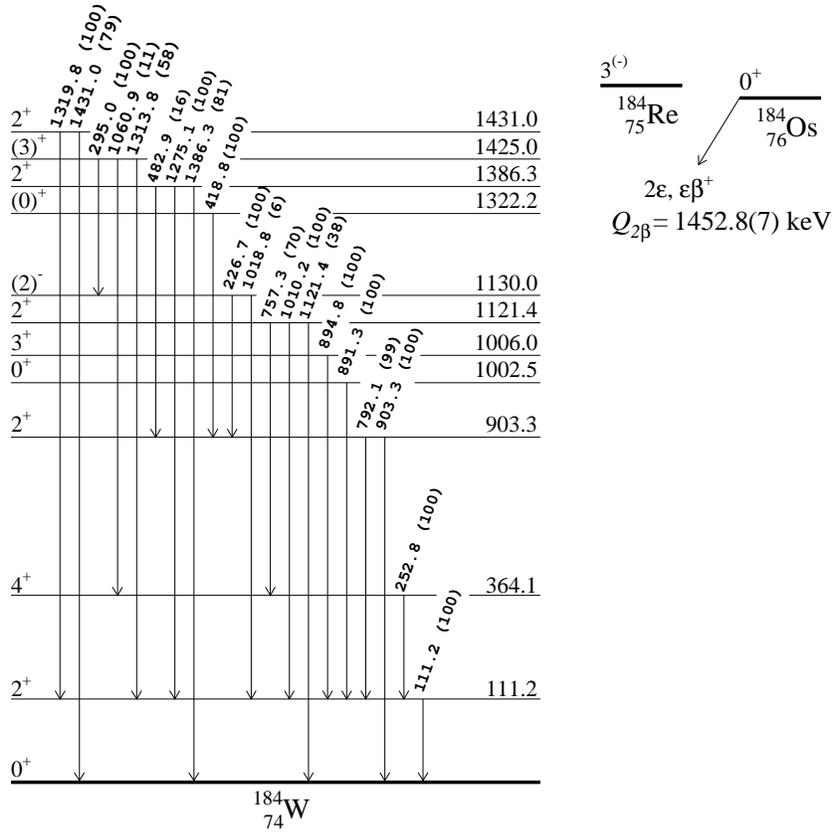,height=11.0cm}}
\caption{A simplified decay scheme of $^{184}$Os \cite{NDS-184}.
The $Q_{2\beta}$ value is from \cite{Wang:2017}. The energies of
the excited levels and of the emitted $\gamma$ quanta are in keV.
The relative intensities of $\gamma$ quanta are given in
parentheses.}
 \label{fig:184Os-decay-scheme}
 \end{center}
 \end{figure}

The $^{184}$Os isotope is one of the potentially 2EC and
EC$\beta^+$ decaying nuclides with the decay energy
$Q_{2\beta}=1452.8(7)$ keV \cite{Wang:2017} and a rather small
isotopic abundance $\delta=0.02(2)\%$ \cite{Meija:2016}. A
simplified decay scheme of $^{184}$Os is shown in Fig.
\ref{fig:184Os-decay-scheme}.

\begin{figure}[!ht]
 \begin{center}
 \mbox{\epsfig{figure=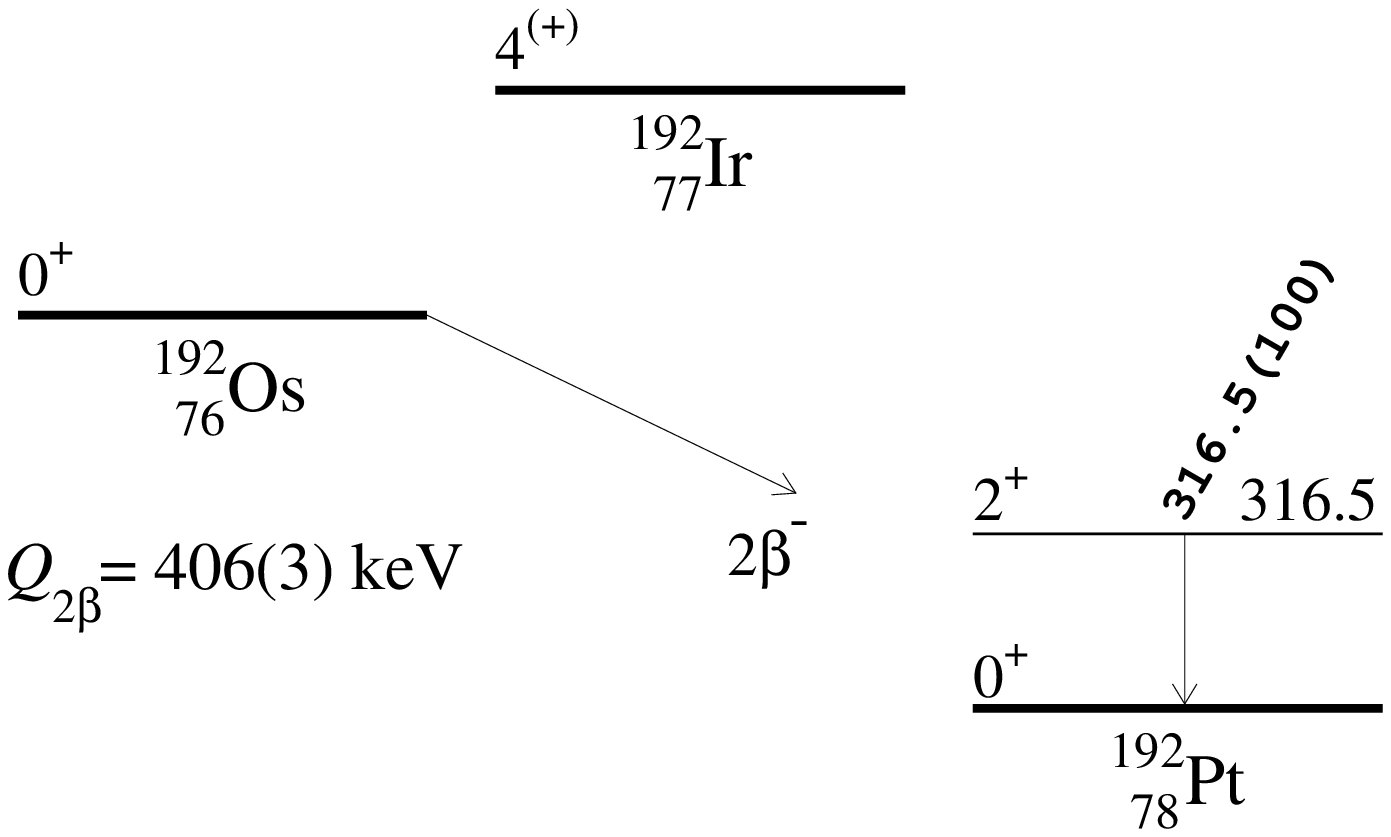,height=4.5cm}}
\caption{Decay scheme of $^{192}$Os \cite{NDS-192}. The
$Q_{2\beta}$ value is from \cite{Wang:2017}.}
 \label{fig:192Os-decay-scheme}
 \end{center}
 \end{figure}

Another osmium isotope, the $^{192}$Os, is potentially $2\beta^-$ active
with decay energy $Q_{2\beta}=406(3)$ keV \cite{Wang:2017} and
isotopic abundance $\delta=40.78(32)\%$ \cite{Meija:2016}. The
decay scheme of $^{192}$Os is shown in Fig.
\ref{fig:192Os-decay-scheme}. As one can see from Figs.
\ref{fig:184Os-decay-scheme} and \ref{fig:192Os-decay-scheme},
some possible $2\beta$ decay processes in $^{184}$Os and
$^{192}$Os are expected to be accompanied by emission of $\gamma$
quanta that can be identified by the $\gamma$-spectrometry method.

An experiment to search for rare decays of naturally occurring
osmium nuclides with emission of $\gamma$ quanta has been carried
out at the LNGS (depth of 3600 meters of water equivalent (m
w.e.)) by using a sample of ultrapure osmium and
ultra-low-background germanium $\gamma$ detectors of the STELLA
facility \cite{STELLA}. The data of the experiment's first stage
\cite{Belli:2012,Belli:2013} were used to derive half-life limits
at the level of $\lim T_{1/2} \sim 10^{14}-10^{17}$ yr for 2EC and
EC$\beta^+$ processes in $^{184}$Os, and the $\lim T_{1/2}=
5.3\times10^{19}$ yr was reached for the $2\beta^-$ decay of
$^{192}$Os to the first $2^+$ 316.5 keV excited level of
$^{192}$Pt. However, the experiment's sensitivity was limited by a
rather poor detection efficiency due to self-absorbtion of
$\gamma$ quanta in the osmium sample that was in the form of four
metal ingots with diameter in the range of (7--10) mm \footnote{It
should be noted that osmium is the densest naturally occurring
element with the density of 22.587 g/cm$^3$ \cite{Haynes:2017}.}.

The poor detection efficiency was especially a troublesome issue
for the main goal of the experiment, which was the search
for $\alpha$ decay of $^{184}$Os and $^{186}$Os to excited levels
of daughter nuclei with emission of $\gamma$ quanta of rather low
energies 103.6 keV and 100.1 keV, respectively. To increase the
detection efficiency, the ingots were cut into thin slices with a
thickness of $(0.79-1.25)$ mm by using a method of electroerosion
cutting. Moreover, the Os sample was placed on a specially
developed ultra-low background broad-energy germanium (BEGe)
detector with improved detection efficiencies and energy
resolutions in the energy region from several keV to several hundreds
keV. The optimization exhibited a substantial improvement of
the experimental sensitivity and lower limits were set for the
$\alpha$ decays of $^{184}$Os and $^{186}$Os to the first excited
levels of daughter nuclei as $\lim
T_{1/2}(^{184}$Os$)=6.8\times10^{15}$~yr and $\lim
T_{1/2}(^{186}$Os$)=3.3\times10^{17}$~yr \cite{Belli:2020}. It
should be noted that the limits are already well above the
theoretical predictions for the nuclides half-lives relative to
the $\alpha$ decays \cite{Belli:2020}.

A new stage of the experiment is in progress by using the Os
sample directly placed on the crystal of the Ge detector  as described in ref. \cite{xx}
to further increase
the detection efficiency of the low-energy $\gamma$-ray quanta
expected in the $\alpha$ decays of $^{184}$Os and $^{186}$Os to
the first excited levels of the daughter nuclei. On the other token,
in the present study the data of
the already completed measurements \cite{Belli:2020} are utilized
to search for $2\beta$ processes in
$^{184}$Os and $^{192}$Os with emission of X-ray and $\gamma$-ray
quanta.

\section{Experiment and data analysis}
\label{sec:exp}

The production of the Os sample, the accurate determination of the
Os sample isotopic composition, the low-background measurements
and data analysis are described in detail in \cite{Belli:2020}.
Here we briefly describe the main features of the experiment.

The Os sample was prepared from osmium metal obtained by
electron-beam melting with further purification by electron-beam
zone refining. The obtained ingots were cut (after the first stage
of the experiment \cite{Belli:2012}) into thin slices with a
thickness of $(0.79-1.25)$ mm by electroerosion cutting with a brass
wire in kerosene.

The isotopic composition of the osmium sample was measured by
using negative thermal ionization mass spectrometry. The isotopic
concentration of $^{184}$Os was determined as 0.0170(7)\%
\cite{Belli:2020}, with an accuracy essentially higher than that
of the adopted reference value of 0.02(2)\% \cite{Meija:2016}. The
$^{192}$Os isotopic abundance was measured as 40.86(5)\% (the
reference value is 40.78(32)\% \cite{Meija:2016}).

The Os slices with a total mass of 117.96(2) g were assembled on
the top and around of the BEGe detector endcap. The active volume of the
detector is 112.5 cm$^3$; the endcap of the detector is made of
1.5 mm aluminium to increase the detection efficiency to low
energy $\gamma$ quanta. The detector with the Os sample was
shielded by layers of $\approx5$ cm thick high-purity copper and
20 cm thick lead. The experiment was running at the STELLA
facility of the LNGS.

 \begin{figure}[!ht]
 \begin{center}
 \mbox{\epsfig{figure=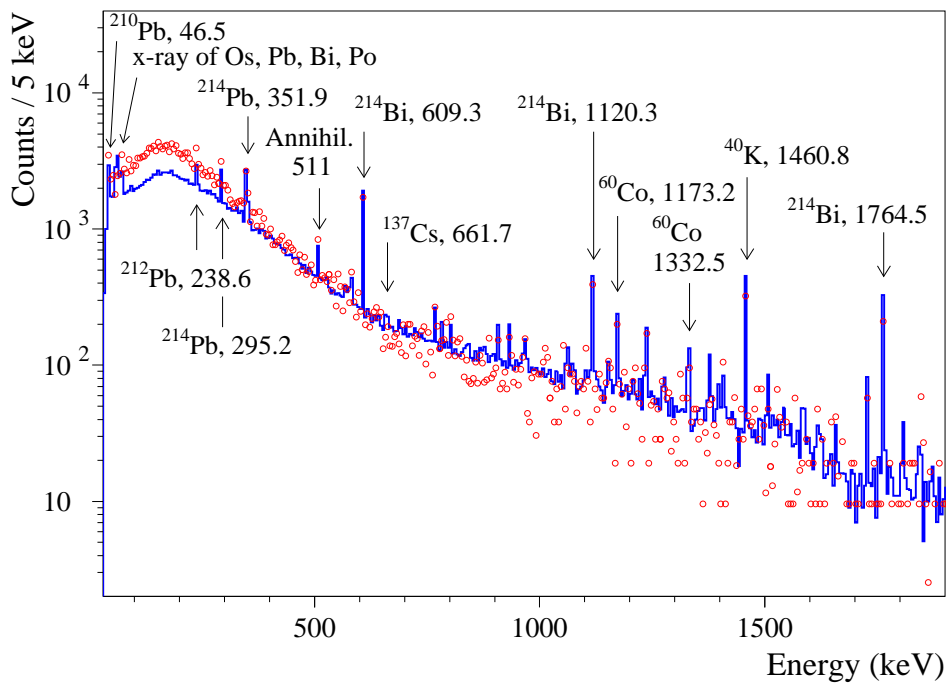,height=9.0cm}}
\caption{Energy spectrum measured with the Os sample by the
ultra-low-background BEGe detector over 15851 h (solid line). The detector's background energy spectrum over 1660 h, normalized to the Os sample's measured time, is shown by circles.
The energies of $\gamma$ peaks are in
keV.}
 \label{fig:Os-BG}
 \end{center}
 \end{figure}

The energy spectra measured with the Os sample over 15851 h and
background data taken over 1660 h are presented in Fig.
\ref{fig:Os-BG}. The lower counting-rate in the spectrum with the
sample below $\sim0.4$ MeV can be explained by the very high
density of osmium that absorbs effectively external radiations
from the shielding materials surrounding the detector (mainly
bremsstrahlung from $^{210}$Bi in the lead shield). The analysis of
the $\gamma$ peaks observed in the spectra allowed us to estimate the
Os sample radioactive contamination to be rather low, with only
$^{40}$K and $^{137}$Cs detected (see Table \ref{tab:rad-cont} and
\cite{Belli:2012} for details of the analysis). Other $\gamma$
peaks observed in the spectrum measured with the Os sample are
statistically indistinguishable from the background. Thus only
limits were set for $^{60}$Co and daughters of $^{232}$Th,
$^{235}$U and $^{238}$U (a possible contamination of the sample by
$^{241}$Am will be discussed in Section \ref{sec:184os}).


\begin{table}[ht]
 \caption{Radioactive trace impurities in the Os
sample.}
\begin{center}
\begin{tabular}{|l|l|l|}
 \hline
 Decay chain    & Radionuclide  & Specific activity (mBq/kg) \\
 \hline
 ~              & $^{40}$K      & $11\pm4$ \\
 ~              & $^{60}$Co     & $\leq1.3$ \\
 ~              & $^{137}$Cs    & $0.5\pm0.1$ \\
 ~              & $^{241}$Am    & $\leq5.6$ \\
 \hline
 $^{232}$Th     & $^{228}$Ra    & $\leq6.6$ \\
 ~              & $^{228}$Th    & $\leq16$ \\
 \hline
 $^{235}$U      & $^{235}$U     & $\leq8.0$ \\
 ~              & $^{231}$Pa    & $\leq3.5$ \\
 ~              & $^{227}$Ac    & $\leq1.1$ \\
 \hline
 $^{238}$U      & $^{238}$U     & $\leq35$ \\
 ~              & $^{226}$Ra    & $\leq4.4$ \\
 ~              & $^{210}$Pb    & $\leq180$ \\
\hline
\end{tabular}
 \label{tab:rad-cont}
\end{center}
\end{table}

The dependence of the energy resolution of the detector (full width at
half maximum, FWHM, in keV) on the energy of $\gamma$-ray quanta
($E_{\gamma}$, in keV) was determined using the $\gamma$
peaks of $^{40}$K, $^{208}$Tl, $^{210}$Pb, $^{214}$Pb, and
$^{214}$Bi observed in the data of the long-time measurements as
following:

\begin{equation}
\rm{FWHM~(keV)}=0.57(5)+0.029(2)\times\sqrt{\textit{E}_{\gamma}}.
\label{eq:fwhm}
\end{equation}

\subsection{Limits on 2EC and EC$\beta^+$ processes in $^{184}$Os}
 \label{sec:184os}

There are no peculiarities in the experimental data which could be
ascribed to $2\beta$-decay processes in $^{184}$Os. Thus, we
report the limits on the $^{184}$Os half-life relative to the
different channels and modes of the $2\beta$ decay adopting the
following formula:

\begin{equation}
 \lim T_{1/2}=\frac{N\cdot \ln 2\cdot \eta \cdot  t}{\lim S},
 \label{eq:t1/2}
\end{equation}

\noindent where \textit{N} is the number of $^{184}$Os nuclei in
the sample ($6.35\times10^{19}$), $\eta$~is the detection
efficiency, $t$ is the time of measurement, and $\lim S$ is the
upper limit on the number of events of the effect searched for
which can be excluded at a given confidence level (C.L.).

The fastest decay of $^{184}$Os is theoretically expected to be the
$2\nu$2EC, mainly absorbing
the K and/or L electron shells. In case of $2\nu$2K, $2\nu$KL,
and $2\nu$2L capture in $^{184}$Os, a cascade of X-rays and Auger
electrons of the W atom is expected. However, the $2\nu$2L decay
cannot be detected in the present experiment since the energies of
the L X-rays of tungsten (7.4 keV--11.7 keV) are below the
detector's energy threshold.

\begin{figure}[!ht]
 \begin{center}
 \mbox{\epsfig{figure=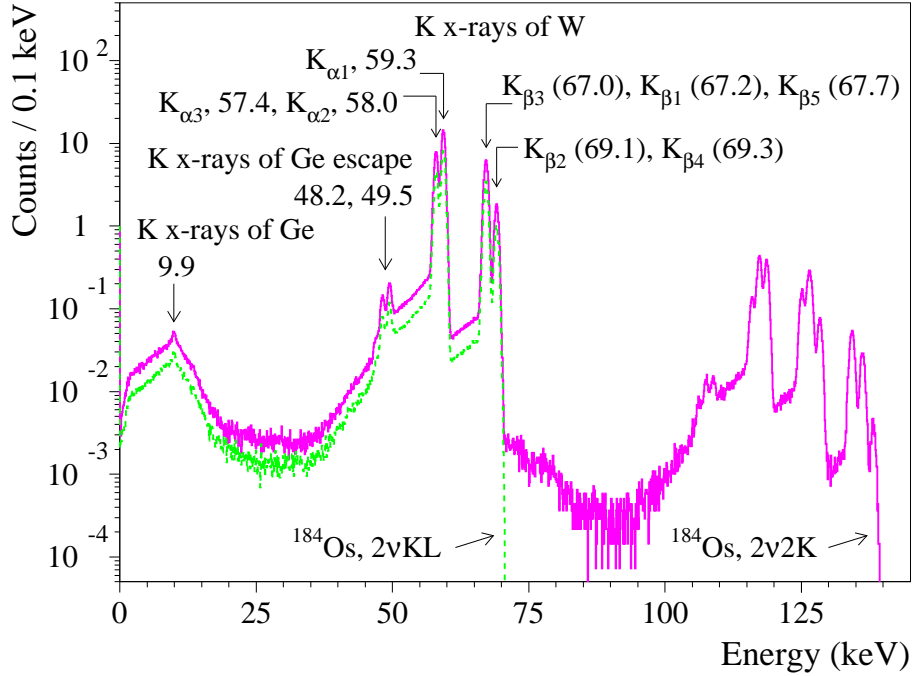,height=9.0cm}}
\caption{Simulated energy spectra for the $2\nu$2K and $2\nu$KL decays of $^{184}$Os assuming the experimental configuration described in Section 2.
Both the distributions are normalized to
$10^4$ decays in the Os sample.}
 \label{fig:2n2Ksim}
 \end{center}
 \end{figure}

The response of the BEGe detector to the $2\nu$2K and $2\nu$KL decays
of $^{184}$Os was simulated with the help of the EGSnrc package
\cite{EGS} assuming the following energies and intensities of
X-ray from the K shell of W atom: 57.43 keV (K$_{\alpha
3}$, 0.021\%), 57.98 keV (K$_{\alpha 2}$, 27.4\%), 59.32 keV
(K$_{\alpha 1}$, 47.6\%), 66.95 keV (K$_{\beta 3}$, 5.33\%), 67.24
keV (K$_{\beta 1}$, 10.3\%), 67.69 keV (K$_{\beta 5}$, 0.24\%),
69.07 keV (K$_{\beta 2}$, 3.56\%), 69.27 keV (K$_{\beta 4}$,
0.54\%) \cite{TOI}. For L$_1$, L$_2$ and L$_3$ shells X-ray quanta
with the mean energy value of 9.5 keV and intensity of 25\% were
simulated. Auger electrons and X-rays with smaller energy were not simulated,
considering the very low probability of reaching the detector for such small
energy electrons and X-ray quanta. The
simulated energy distributions for the $2\nu$2K and $2\nu$KL
decays of $^{184}$Os are shown in Fig. \ref{fig:2n2Ksim}.

\begin{figure}[!ht]
 \begin{center}
 \mbox{\epsfig{figure=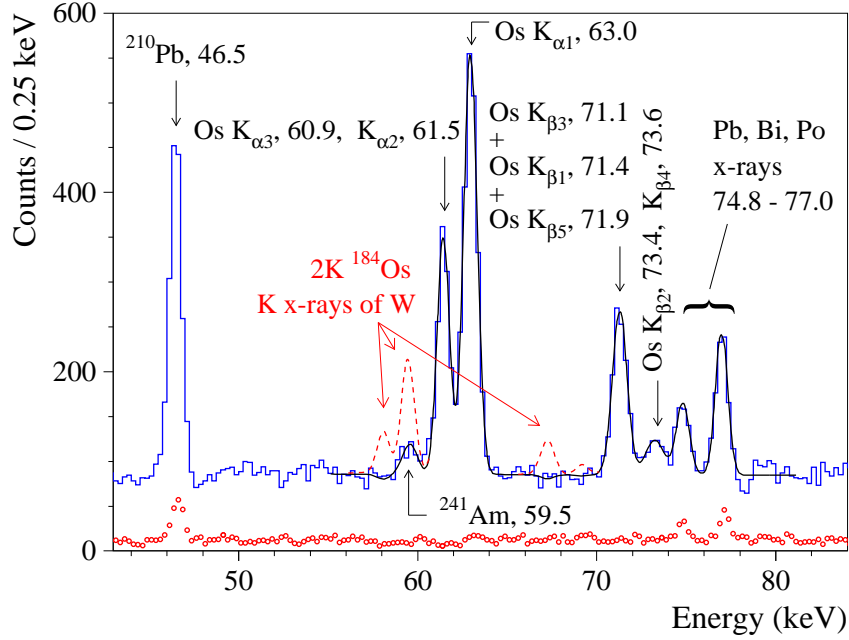,height=8.5cm}}
\caption{Energy spectra measured with the Os sample (solid
histogram) and background data (circles) in the energy region
where K X-rays of W are expected from the $2\nu$2K and $2\nu$KL
decays of $^{184}$Os. The fit of the data by the background model
is shown by solid line, while the excluded effect of the $2\nu$2K
decay is presented by dashed line (the excluded distribution is
multiplied by a factor 10 for better visibility). Note that the two
spectra are not normalized to the time;
the acquisition times are 15851 h and 1660 h, respectively.}
 \label{fig:2n2K}
 \end{center}
 \end{figure}

The energy spectrum accumulated with the Os sample was fitted by
the simulated $2\nu$2K ($2\nu$KL) distribution plus a sum of
several Gaussian functions to describe the X-ray peaks of Os, Pb,
Bi and Po present in the spectrum (see Fig.
\ref{fig:2n2K})\footnote{It should be noted that K X-ray peaks of
Os are absent in the background data, while the Pb, Bi and Po
X-ray peaks are clearly observed in both spectra.}.
The fits in the energy interval $(55-81)$ keV, where K X-rays of W are,
return area of the effects searched for: $S=(-74\pm86)$ counts
($(-87\pm83)$ counts) that is no evidence for the $2\nu$2K ($2\nu$KL)
decay. The fits quality is very good:
$\chi^2/$n.d.f.$=61.5/81=0.759$ ($\chi^2/$n.d.f.$=59.2/81=0.731$),
where n.d.f. is number of degrees of freedom. According to the
recommendations \cite{feld98} 78 (65) counts should be accepted as
$\lim S$ at 90\% C.L.\footnote{All the half-life limits in the
present work are given at 90\% C.L.} Taking into account the
detection efficiency simulated with the help of the EGSnrc package
$\eta=2.911\%$ (1.635\%), the following half-life limits were set
for the $2\nu$2K and $2\nu$KL decays of $^{184}$Os to the ground
state of $^{184}$W:

\begin{center}

$T_{1/2}(2\nu\rm{2K})>3.0\times10^{16}$~yr,

\vskip 0.3cm

$T_{1/2}(2\nu\rm{KL})>2.0\times10^{16}$~yr.

\end{center}

The energies of X-ray quanta used in the analysis, the detection
efficiencies, the value of $\lim S$, the half-life limits obtained
in the present study and the ones obtained in the previous work
for the $2\nu$2K decay of $^{184}$Os \cite{Belli:2012} are given
in Table \ref{tab:results}.

 \begin{table*}[htbp]
  \vspace{-0.7cm}
 \caption{Half-life limits on
2$\beta$ processes in $^{184}$Os and $^{192}$Os.}
\begin{center}
\resizebox{1.25\textwidth}{!}{
\begin{tabular}{|l|l|l|l|l|l|l|}
\hline
 Transition                 & Level of          & $E_\gamma$        & Detection     & $\lim S$  & \multicolumn{2}{c|}{Experimental limits,} \\
 ~                          & daughter          & (keV)             & efficiency    & ~         & \multicolumn{2}{c|}{$T_{1/2}$ (yr) at 90\% C.L.} \\
 \cline{6-7}
 ~                          & nucleus           &                   & ~             &  ~        & Present work              & Previous result  \\
 ~                          & (keV)             &                   & ~             &  ~        &           ~               & \cite{Belli:2012} \\
 \hline
 $^{184}$Os$~\to~$$^{184}$W  & ~                 & ~                 & ~             & ~         &           ~               & ~ \\
 $2\nu$2K                   & g.s.              & $57-69$           & 2.911\%       & 78        & $\geq3.0\times10^{16}$    & $\geq1.9\times10^{14}$ \\
 $2\nu$KL                   & g.s.              & $57-69$           & 1.635\%       & 65        & $\geq2.0\times10^{16}$    & ~ \\
 $2\nu$2K                   & $2^{+}~111.2$     & $57-69$           & 3.487\%       & 78        & $\geq3.6\times10^{16}$    & $\geq3.1\times10^{15}$  \\
 $2\nu$KL                   & $2^{+}~111.2$     & $57-69$           & 1.959\%       & 65        & $\geq2.4\times10^{16}$    & $\geq3.1\times10^{15}$  \\
 $2\nu$2EC                  & $2^{+}~111.2$     & 111.2             & 0.340\%       & 37        & $\geq7.3\times10^{15}$    & $\geq3.1\times10^{15}$ \\
 $2\nu$2EC                  & $2^{+}~903.3$     & 903.3             & 1.230\%       & 4.9       & $\geq2.0\times10^{17}$    & $\geq3.2\times10^{16}$ \\
 $2\nu$2EC                  & $0^{+}~1002.5$    & 891.3             & 2.397\%       & 6.8       & $\geq2.8\times10^{17}$    & $\geq3.8\times10^{17}$ \\
 $2\nu$2EC                  & $2^{+}~1121.4$    & 757.3             & 0.802\%       & 6.2       & $\geq1.0\times10^{17}$    & $\geq6.9\times10^{16}$ \\
 $2\nu$KL                   & $(0^{+})~1322.2$  & 903.3             & 1.056\%       & 4.9       & $\geq1.7\times10^{17}$    & ~ \\
 $2\nu$2L                   & $2^{+}~1386.3$    & 1275.1            & 0.967\%       & 26        & $\geq3.0\times10^{16}$    & ~ \\
 $2\nu$2L                   & $(3)^{+}~1425.0$  & 903.3             & 0.518\%       & 4.9       & $\geq8.4\times10^{16}$    & ~ \\
 $2\nu$2L                   & $2^{+}~1431.0$    & 1319.8            & 1.002\%       & 18        & $\geq4.4\times10^{16}$    & ~ \\
 ~                          & ~                 & ~                 & ~             & ~         &                           &  ~                     \\
 $0\nu$2K                   & g.s.              & $1313.1-1314.5$   & 1.838\%       & 9.0       & $\geq1.6\times10^{17}$    & $\geq2.0\times10^{17}$ \\
 $0\nu$KL                   & g.s.              & $1370.5-1373.8$   & 1.827\%       & 11        & $\geq1.3\times10^{17}$    & $\geq1.3\times10^{17}$ \\
 $0\nu$2L                   & g.s.              & $1427.9-1433.1$   & 1.833\%       & 20        & $\geq7.3\times10^{16}$    & $\geq1.4\times10^{17}$ \\
 $0\nu$2K                   & $2^+$  111.2      & $1201.9-1203.3$   & 1.911\%       & 20        & $\geq7.6\times10^{16}$    & $\geq3.3\times10^{17}$ \\
 $0\nu$KL                   & $2^+$  111.2      & $57-69$           & 1.584\%       & 65        & $\geq1.9\times10^{16}$    &  \\
 $0\nu$2EC                  & $2^+$  903.3      & 903.3             & 1.019\%       & 4.9       & $\geq1.7\times10^{17}$    & $\geq2.8\times10^{16}$ \\
 $0\nu$2EC                  & $0^+$  1002.5     & $310.6-312.0$     & 3.773\%       & 14        & $\geq2.1\times10^{17}$    & $\geq3.5\times10^{17}$ \\
 $0\nu$2EC                  & $2^+$ 1121.4      & 757.3             & 0.736\%       & 6.2       & $\geq9.4\times10^{16}$    & $\geq6.4\times10^{16}$ \\
 $0\nu$KL                   & $(0)^+$ 1322.2    & 903.3             & 1.045\%       & 4.9       & $\geq1.7\times10^{17}$    & $\geq2.8\times10^{16}$ \\
 $0\nu$2L                   & $2^+$ 1386.3      & 1275.1            & 0.966\%       & 26        & $\geq3.0\times10^{16}$    & $\geq6.7\times10^{16}$ \\
 $0\nu$2L                   & $(3)^{+}~1425.0$  & 903.3             & 0.517\%       & 4.9       & $\geq8.4\times10^{16}$    & ~ \\
 Resonant $0\nu$2L          & $2^+$ 1431.0      & 1319.8            & 1.005\%       & 18        & $\geq4.4\times10^{16}$    & $\geq8.2\times10^{16}$ \\
 ~                          & ~                 & ~                 & ~             & ~         &                           &  ~                     \\

 $2\nu$EC$\beta^+$          & g.s.              & 511               & 7.526\%       & 58        & $\geq1.0\times10^{17}$    & $\geq2.5\times10^{16}$ \\
 $2\nu$EC$\beta^+$          & $2^{+}~111.2$     & 511               & 7.271\%       & 58        & $\geq1.0\times10^{17}$    & $\geq2.5\times10^{16}$ \\

 $0\nu$EC$\beta^+$          & g.s.              & 511               & 7.403\%       & 58        & $\geq1.0\times10^{17}$    & $\geq2.5\times10^{16}$ \\
 $0\nu$EC$\beta^+$          & $2^{+}~111.2$     & 511               & 7.191\%       & 58        & $\geq9.9\times10^{16}$    & $\geq2.4\times10^{16}$ \\

 ~                          & ~                 & ~                 & ~             & ~         &                           &  ~                     \\
 $^{192}$Os$~\to~^{192}$Pt   & ~                 & ~                 & ~             & ~         &                           &  ~                     \\
 $2\beta^{-}$ ($2\nu+0\nu$) & $2^+$ 316.5       & 316.5             & 4.820\%       & 45        &  $\geq2.0\times10^{20}$   & $\geq5.3\times10^{19}$  \\
 \hline
\end{tabular}
}
\end{center}
 \label{tab:results}
\end{table*}


Every 2EC transitions of $^{184}$Os to excited levels of $^{184}$W
should be accompanied by X-rays of W, too. Thus, the half-life
limits on the $2\nu$2K and $2\nu$KL decays of $^{184}$Os to the
111.2 keV excited level of $^{184}$W were estimated by using the
already obtained $\lim S$ values for the $2\nu$2K and $2\nu$KL
decays to the ground state of $^{184}$W. The detection
efficiencies for the decays calculated with the EGSnrc package
and the event generator DECAY0 \cite{DECAY0a,DECAY0b} are slightly
higher\footnote{This happens due to a rather big internal electron
conversion coefficient $\alpha=2.57$ for the $\gamma$ quanta
emitted in de-excitation of the 111.2 keV excited level of
$^{184}$W. As a result, additional K X-rays are coming from the
internal conversion process of the 111.2-keV level de-excitation.}
than those relative to the g.s. transitions, leading to
a slightly higher half-life limits on the decays (the
limits are presented in Table \ref{tab:results}).

 \begin{figure}[!ht]
 \begin{center}
 \mbox{\epsfig{figure=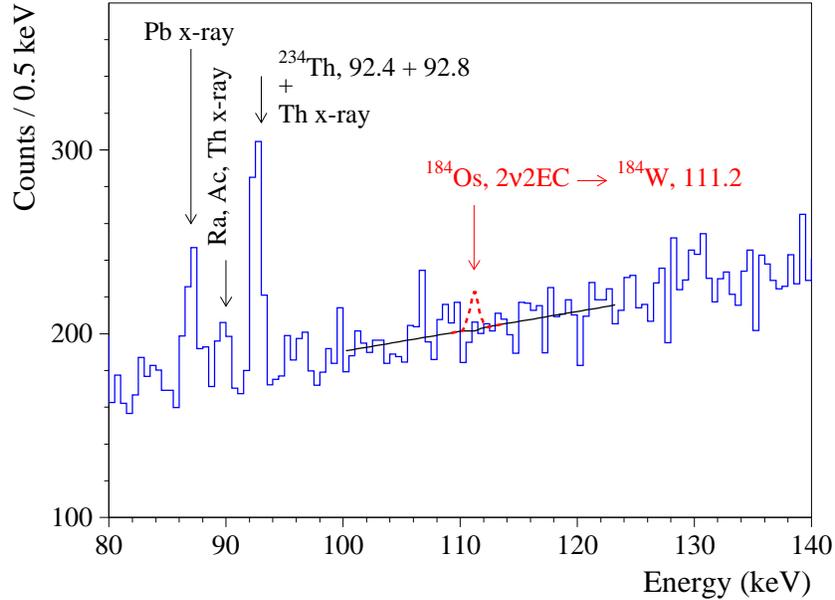,height=8.0cm}}
\caption{Part of the energy spectrum measured with the Os sample
near the 111.2 keV $\gamma$ peak expected in the
$2\nu$2EC decay of $^{184}$Os to the 111.2 keV excited level of
$^{184}$W. The fit of the data is shown by the solid line, the
excluded peak is shown by the dashed line.}
 \label{fig:111}
 \end{center}
 \end{figure}


Another approach to search for 2EC transitions of $^{184}$Os to
excited levels of $^{184}$W is based on the analysis of the
experimental data in a region where $\gamma$ peaks after
de-excitation of $^{184}$W are expected\footnote{It should be
stressed that limits obtained in such a way are valid for capture
from any shells, not only for 2K or KL decays.}. For example, Fig.
\ref{fig:111} shows a part of the energy spectrum measured with
the Os sample where a $\gamma$ peak with the energy 111.2 keV
after the $2\nu$2EC decay of $^{184}$Os to the 111.2 keV excited
level of $^{184}$W is expected. The data were fitted by a simple
model consisting in a linear function (to describe the background)
and a Gaussian function centered at 111.2 keV with the width
calculated by applying the formula (\ref{eq:fwhm}) as the effect
searched for. The fit returns a peak area $(-2.1\pm23.7)$ counts
that gives $\lim S=37$ counts at 90\% C.L. (the excluded peak is
shown in Fig. \ref{fig:111}, too). The detection efficiency for
$\gamma$ quanta with energy 111.2 keV in the decay was simulated
by using the EGSnrc package and the event generator DECAY0
obtaining $\eta=0.340\%$. This substantially smaller detection
efficiency (in comparison to the X-ray expected in the $2\nu$2K
and $2\nu$KL decays) can be explained by the rather big internal
electron conversion coefficient $\alpha=2.57$ for the 111.2-keV
$\gamma$-transition. As a result, the half-life limit for the
$2\nu$2EC decay of $^{184}$Os to the 111.2 keV excited level of
$^{184}$W is $\lim T_{1/2}=7.3\times10^{15}$~yr.

The half-life limits for the decays of $^{184}$Os to several excited
levels of $^{184}$W were obtained in a similar way. The results of
the analysis are presented in Table \ref{tab:results}.

\begin{figure}[!ht]
 \begin{center}
 \mbox{\epsfig{figure=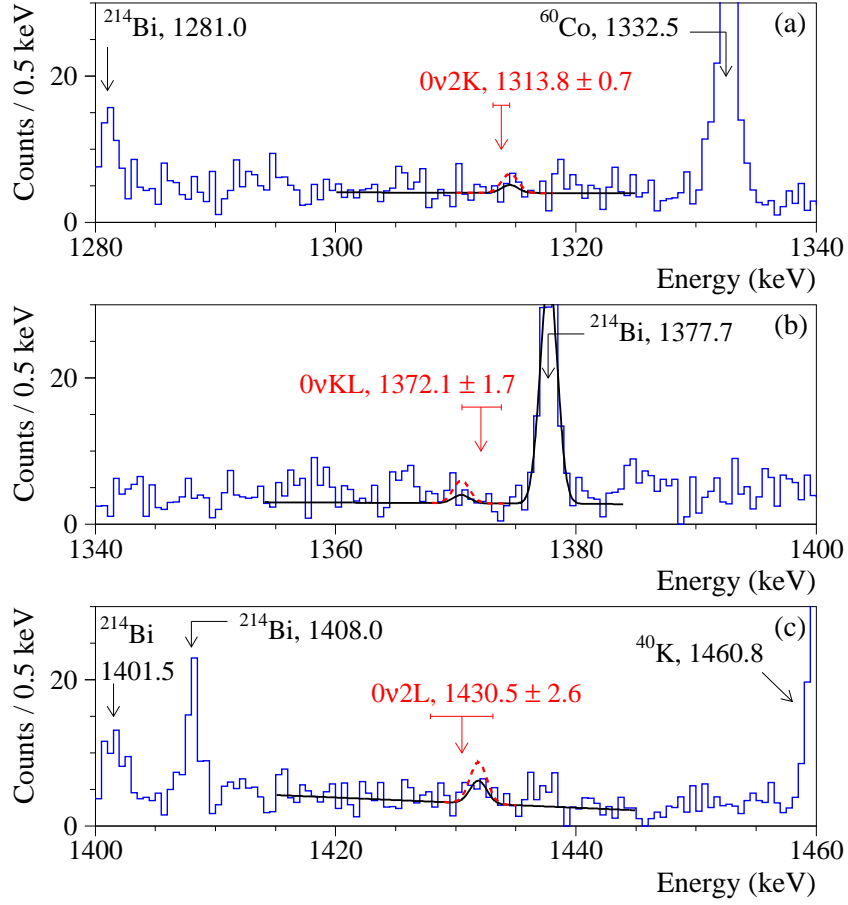,height=12.0cm}}
\caption{Parts of the energy spectrum measured with the Os sample
where bremsstrahlung $\gamma$ peaks from the $0\nu$2K (a),
$0\nu$KL (b), and $0\nu$2L (c) captures in $^{184}$Os to the
ground state of $^{184}$W are expected. The fits of the data are
shown by solid lines, while the excluded peaks are presented by
dashed lines. The horizontal lines (above the arrows labelling the
energy of the peaks searched for) show the energy uncertainty due
the error of the $Q_{2\beta}$ value of $^{184}$Os plus the
difference of the L$_1$, L$_2$ and L$_3$ shells binding energies.
}
 \label{fig:0n2e}
 \end{center}
 \end{figure}

In the neutrinoless 2EC process in $^{184}$Os from K and L shells
to the ground state of $^{184}$W the energy excess is supposed to
be taken away by a bremsstrahlung $\gamma$ quanta with an energy
$E_{\gamma}=Q_{2\beta}-E_{b1}-E_{b2}$, where $E_{bi}$ are the
binding energies of the captured electrons on the atomic shells of
the daughter W nuclide. For example, to estimate the half-life
limit for the $0\nu$2K decay of $^{184}$Os, the experimental
spectrum measured with the Os sample was fitted in the energy
interval $1300-1325$ keV by the sum of a straight line
(background) and a Gaussian-shaped peak with a width determined
using the formula (\ref{eq:fwhm}). The peak position was varied
within the energy interval $(1313.8\pm0.7)$ keV determined by the
accuracy of the $Q_{2\beta}$ value \cite{Wang:2017}. The maximum
peak area returned by the fits $(2.3 \pm 4.1)$ counts, see Fig.
\ref{fig:0n2e} (a)) was considered, leading to $\lim S=9.0$ counts
and $\lim T_{1/2}=1.6\times 10^{17}$ yr. In the case of the
$0\nu$KL and $0\nu$2L decays the ranges of the intervals in which
the peaks searched for may appear were slightly larger due to the
different values of the binding energies on the L$_1$, L$_2$ and
L$_3$ shells of the W atom: 12.1 keV, 11.5 keV and 10.2 keV,
respectively. The parts of the energy spectrum measured with the
Os sample in the regions of interest, the fitting curves, and the
excluded peaks are shown in Fig. \ref{fig:0n2e}.

The highest sensitivity to the $0\nu$2EC decays of $^{184}$Os to
the excited levels of $^{184}$W (denoted here as $E^{*}$) was
achieved either searching for $\gamma$ peaks with certain energy
emitted after de-excitation of the excited levels (as e.g. for the
transitions to the levels $2^+$ 903.3 keV and $2^+$ 1121.4 keV) or
analysing the data in the energy region where the
bremsstrahlung $\gamma$ quanta with the energy
$E_{\gamma}=Q_{2\beta}-E^{*}-E_{b1}-E_{b2}$ are expected (the case
of decays to the levels $2^+$ 111.2 keV and $0^+$ 1002.5 keV). The
results of the analysis are again presented in Table
\ref{tab:results}.

Neutrinoless double-electron captures to the excited levels
$(0)^+$ 1322.2 keV and $2^+$ 1386.3 keV have been considered as
resonant ones \cite{Belli:2012}. However, double-electron capture
to the excited level $(0)^+$ 1322.2 keV is energetically favoured
from K and L shells\footnote{Although with a lower probability,
capture from higher shells (M, N, O) is also allowed.}, while the
$2^+$ 1386.3 keV level favours the electron capture from the  L
shell. The transitions are far from resonant condition taking into
account the rather big difference $\Delta =
Q_{2\beta}-E^{*}-E_{b1}-E_{b2}$: $\Delta =[(49.0-50.9)\pm0.7]$ keV
for the $0\nu$KL to the 1322.2-keV level, and $\Delta
=[(42.3-46.1)\pm0.7]$ keV for the $0\nu$2L transition to the
1386.3-keV level. Nevertheless, the decays were restricted
analysing the experimental data in the energy regions where
intense $\gamma$ peaks from the decays are expected. In addition,
a half-life limit on the kinematically allowed $0\nu$2L transition
of $^{184}$Os to the $(3)^{+}~1425.0$ excited level of $^{184}$W
with $\Delta =[(3.6-7.4)\pm0.7]$ keV was set in the present work
(the results of the analysis are presented in Table
\ref{tab:results}). Only the transition to the $2^+$ 1431.0 keV
level from the L shells remains a candidate for the resonant decay
taking into account the region of $\Delta$ values from -3.1 keV to
2.1 keV. A search for the resonant $0\nu$2L transition was
realized analysing the experimental energy spectrum gathered with
the Os sample in the energy region where a peak with energy 1319.8
keV is expected. However, the obtained limit (see Table
\ref{tab:results}) is weaker than the previous result due to the
rather low detection efficiency for high energy $\gamma$-ray
quanta of the used BEGe detector.

\begin{figure}[!ht]
 \begin{center}
 \mbox{\epsfig{figure=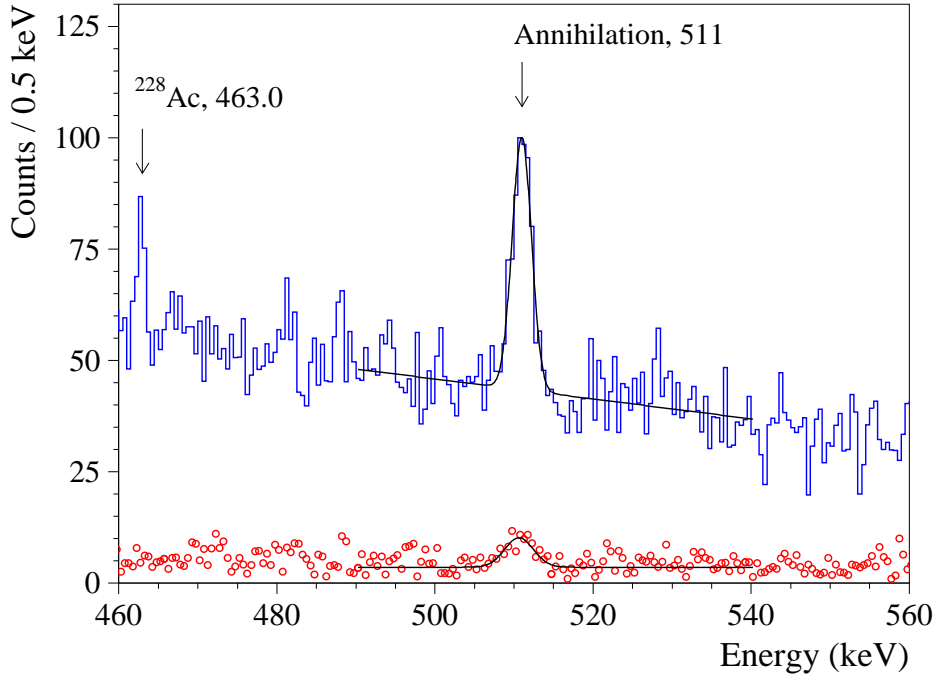,height=9.0cm}}
\caption{Part of the energy spectra measured with the Os sample in
the vicinity of the 511 keV annihilation peak. The background data
are presented by dots. The fits of the data are shown by the solid
lines. Note that the two spectra are not normalized to the time;
the acquisition times are 15851 h and 1660 h, respectively.}
 \label{fig:511}
 \end{center}
 \end{figure}

The electron capture with positron emission in $^{184}$Os should
lead to the emission of two annihilation $\gamma$ quanta with
energy 511 keV, causing an extra counting rate in the annihilation
peak. The energy spectra taken with the Os sample and the
background data were fitted in the energy interval $(490-540)$ keV
with a model constructed from a 511 keV peak and a linear function
to describe background. The peak's width was a free parameter of
the fit to take into account a typically bigger width of the
annihilation peak due to the Doppler broadening. The fits of the
spectra measured with the Os sample and of the background data in
the vicinity of the annihilation peak are shown in Fig.
\ref{fig:511}. There are $(346\pm32)$ counts in the 511 keV peak
in the data gathered with the Os sample, while the
annihilation-peak area in the background spectrum is $(60\pm13)$
counts, which leads to the residual peak area $(-227\pm128)$
counts and to the $\lim S=58$ counts after the normalization of
the expected number of background counts to the time of
measurement with the Os sample. Taking into account the detection
efficiency for the annihilation $\gamma$ quanta (slightly
different depending on the decay mode: $2\nu$ or $0\nu$, see Table
\ref{tab:results}) one can obtain the half-life limits for the
$2\nu$EC$\beta^+$ and $0\nu$EC$\beta^+$ decays of $^{184}$Os to
the ground state of $^{184}$W presented in Table
\ref{tab:results}. Moreover, limits on the $2\nu$EC$\beta^+$ and
$0\nu$EC$\beta^+$ decays to the 111.2 keV excited level of
$^{184}$W were obtained by using the annihilation-peak analysis,
too.

Most of the limits obtained in the present work are higher than
the limits resulted in the previous stage of the experiment
\cite{Belli:2012}. The largest improvement (more than two orders
of magnitude) was achieved for the $2\nu$2K decay, while the
$2\nu$KL process was investigated for the first time. Also the
$2\nu$ double-electron capture to the $(0)^+$ 1322.2 keV, $2^+$
1386.3 keV and $2^+$ 1431.4 keV excited levels of $^{184}$W were
studied for the first time. However, the sensitivity of the
present experiment is worse for some decay channels emitting high energy
$\gamma$ quanta because of the lower detection efficiency of the used BEGe detector.

\subsection{Limit on $2\beta^-$ decay of $^{192}$Os to the first excited level of $^{192}$Pt}
 \label{sec:192os}

There is no evidence in the data for a peak at the energy 316.5
\begin{figure}[!b]
 \begin{center}
 \mbox{\epsfig{figure=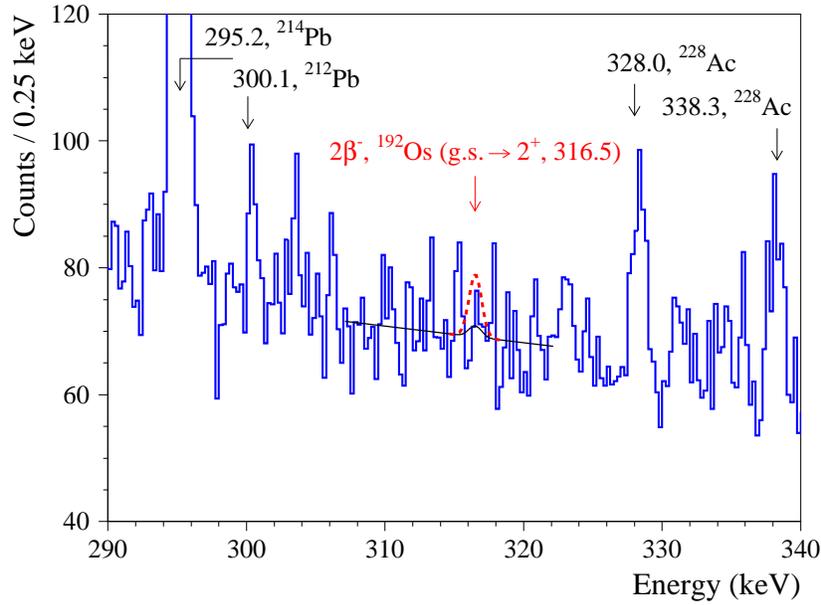,height=8.0cm}}
\caption{Part of the energy spectrum measured with the Os sample
where a peak with energy 316.5 keV from the $2\beta^-$ decay of
$^{192}$Os to the first $2^+$ 316.5 keV excited level of
$^{192}$Pt is expected. The fit of the data is shown by the solid
line, while the excluded $\gamma$ peak is presented by the dashed
line.}
 \label{fig:192Os}
 \end{center}
 \end{figure}
keV, expected in the $2\beta^-$ decay of $^{192}$Os to the first
$2^+$ 316.5 keV excited level of $^{192}$Pt. To estimate a $\lim
S$ value for the 316.5 keV peak area, the energy spectrum taken
with the Os sample was fitted by a model consisting of a Gaussian
peak at energy 316.5 keV (with the width determined by formula
(\ref{eq:fwhm})) and a linear function to describe the background.
The best fit ($\chi^2/$n.d.f.$=31.9/58=0.550$) achieved in the
energy interval ($307-322$) keV returns as peak area: $(8.1 \pm
22.5)$ counts, which corresponds to $\lim S=45$ counts. The
fitting curve and the excluded peak are shown in Fig.
\ref{fig:192Os}. With detection efficiency of $\eta=4.820\%$ and
$1.526\times10^{23}$ number of $^{192}$Os nuclei in the sample,
the half-life limit $T_{1/2}\geq 2.0\times10^{20}$ yr was
obtained, valid for both the $0\nu$ and $2\nu$ decay modes. This
limit improves of 4 times the result of the previous experiment
(see Table \ref{tab:results}).


\section{Conclusions}

Double-beta processes in $^{184}$Os and $^{192}$Os were searched
for over 15851 h using an ultra-low-background
broad-energy germanium detector with an active volume of 112
cm$^3$, optimized for low-energy $\gamma$-ray
spectrometry. The sample of ultra-pure osmium with a mass
of 118 g and with a thickness of $(0.79-1.25)$ mm was placed on
the detector end-cap. The experiment was carried out in the STELLA
facility of the LNGS.

New improved half-life limits on most of the $2\beta$ decay
channels of $^{184}$Os were set at the level of $\lim T_{1/2} \sim
10^{16}-10^{17}$ yr at 90\% C.L. A particular progress was
achieved for the $2\nu$2K and $2\nu$KL processes in $^{184}$Os
(the $2\nu$KL decay was analyzed for the first time) thanks to the
substantial improvement of the detection efficiency with the thin
Os sample and the use of the BEGe detector with high detection
efficiency and good energy resolution to X-ray quanta expected in
the decays. The half-lives of the $^{184}$Os $2\nu$2K and $2\nu$KL
decays were measured to be: $T_{1/2}\geq3.0\times10^{16}$ yr and
$T_{1/2}\geq2.0\times10^{16}$ yr, respectively. The newly
determined lower half-life limit on the $2\beta^-$ decay of
$^{192}$Os to the first excited level of $^{192}$Pt, $\lim
T_{1/2}=2.0\times 10^{20}$ yr, improves the previous limit by 4
times.

The results of the present experiment are very far from
theoretical estimates of the $^{184}$Os decay probability. While
there are no estimates of the half-lives for the two-neutrino 2EC or EC$\beta^+$
processes, the existing calculations for the half-life
of $^{184}$Os concerning the $0\nu$2EC decay to the 1322.2 keV
level of $^{184}$W are at level of $T_{1/2}\sim
10^{28}-10^{31}$ yr (assuming the mechanism of the decay by
exchange of a virtual light Majorana neutrino with $\langle
m_{2\beta}\rangle=0.1$ eV) \cite{Krivoruchenko:2011,Smorra:2012}.

The sensitivity of the experiment could be improved by $4-6$
orders of magnitude by using osmium enriched in the
$^{184}$Os isotope and by increasing the sample mass and the number of
BEGe detectors, and further reducing the background level
(e.g., by utilization of the background-reduction technique
applied in the GERDA experiment \cite{Agostini:2020}).
Surprisingly, such an experiment looks practically realistic
despite the very low natural isotopic abundance of $^{184}$Os.
Osmium isotopes could be enriched by gas centrifugation
\cite{ECP}, at present the only viable technology to produce large
amounts of isotopically enriched materials.

\section{Acknowledgements}

The group from the Institute for Nuclear Research of NASU (Kyiv,
Ukraine) was supported in part by the National Research Foundation
of Ukraine Grant No. 2020.02/0011. D.V.~Kasperovych, and
O.G.~Polischuk were supported in part by the project
``Investigations of rare nuclear processes'' of the program of the
National Academy of Sciences of Ukraine ``Laboratory of young
scientists''.

\end{document}